\documentstyle[twocolumn,prc,aps,psfig]{revtex}
\begin{document}\hbadness=10000
\twocolumn[\hsize\textwidth\columnwidth\hsize\csname %
@twocolumnfalse\endcsname
\title{
Low-\mathversion{bold}$m_\bot$\mathversion{normal} 
\mathversion{bold}$\pi^+$-$\pi^-$\mathversion{normal} Asymmetry
Enhancement  from Hadronization of QGP
}
\author{Jean Letessier and Ahmed Tounsi,}
\address{
Laboratoire de Physique Th\'eorique et Hautes Energies\\
Universit\'e Paris 7, 2 place Jussieu, F--75251 Cedex 05\\
}
\author{and\\ \ \\Johann Rafelski}
\address{
Department of Physics, University of Arizona, Tucson, AZ 85721.
}
\date{November 17, 1999}
\maketitle
\begin{abstract}
We show that in sudden hadronization of  QGP 
a non-equilibrium value of the 
pion phase space occupancy parameter $\gamma_q>1$
is expected in order to accommodate the entropy
excess in QGP and the process of gluon fragmentation. 
When  $\gamma_q$ is near to its maximum allowed value, 
pion overabundance is shown to arise at low $m_\bot$, 
where the charged pion asymmetry is also amplified.
These effects are considered
and we show that their magnitude suffices to
explain pertinent experimental data obtained in 160--200$A$\,GeV  
Pb- and S- induced reactions.
\end{abstract}
]
\begin{narrowtext}
Highly relativistic collisions of heavy nuclei are under 
experimental and theoretical study of which the 
primary aim is the discovery  of the deconfined 
quark-gluon plasma (QGP) phase~\cite{QGP}, and 
the  understanding of its  transformation into hadrons.
We shall demonstrate how the measurement of 
low $m_\bot$  central  rapidity abundance of charged
pion yields, $dN(\pi^-)/dm_\bot$ and $dN(\pi^+)/dm_\bot$, 
provides a sensitive way to probe the outcome of hadronization of QGP. 
We will, in particular, show that the relative deformation
of low-$m_\bot$ pion spectra provides a measure for 
chemical (particle) non-equilibrium overabundance  phenomenon. 

Thermal equilibration ({\it i.e.}, equilibration of the energy distribution)  
of final state hadrons
is  seen in particle spectra to be a well working  hypothesis. This, however,  
does not imply chemical ({\it i.e.}, abundance) equilibration. 
Since chemical hadron abundance equilibrium is approached 
on the same time scale as the duration  of the heavy ion collisions, it has been 
proposed to analyze the  abundances of final state  hadrons allowing
chemical non-equilibrium phase space occupancies \cite{KMR86}.
This has been applied initially only to describe the abundance of 
strange quarks \cite{Raf91}, which
were early on understood to be not necessarily produced rapidly enough \cite{KMR86}. 
More recently, it has been suggested that chemical equilibration 
of light quarks and gluons is indeed not assured at RHIC energies
and beyond \cite{Gei93,Bir93,SMM97}. On the other hand, at SPS, the excess of 
pion multiplicity has been observed and understood to result from 
excess entropy content of the primordial high entropy phase formed in the
interaction \cite{Let93,Let95}. This leads naturally to the hypothesis of
light quark overabundance  generated, {\it e.g.}, by gluon fragmentation
in hadronization of the QGP high entropy phase. This can be most
easily described by counting light valence quark abundance in hadrons 
 \cite{LRa99,LRPb98,RLprl99}, as was done for strangeness.

Indeed, when the multi-particle  production processes, in
158--200$A$ GeV S-- and Pb--Pb collisions carried 
out at CERN-SPS, have been analyzed 
using the methods of the statistical Fermi model \cite{Fer50},
light quark overabundance and the associated chemical 
 non-equilibrium was found. Overall, the results of this analysis
show several features that combined do make a strong case 
that at least the strange hadronic particles seen  at CERN-SPS,
are  emerging nearly directly from sudden hadronization
of deconfined QGP phase of hadronic matter \cite{RLprl99}. 
Moreover, the recognition that the hadron
abundances are better described with oversaturation
of the quark phase space allows a notable reduction of the 
chemical freeze-out temperature, $T_f=145\pm5$\,MeV, accompanied
by a reduction of the baryochemical potential, $\mu_{B}=200$--210\,MeV,
is consistent with the underlying dynamical picture. What remains open 
is the issue if the predominant hadronic fraction, pions, also 
follow the sudden hadronization model, a point we address here.

This general discussion introduced already 
a few statistical parameters of the Fermi model of 
well established meaning and we will complete the list now. Aside of 
the chemical freeze-out temperature $T_{f}$,  further chemical
parameters control hadron abundances: we note the
light quark fugacity $\lambda_u$ and $\lambda_d$\, and the 
phase space occupancy of light quarks 
$\gamma_{q}=\gamma_{u}=\gamma_{d}$.  
We recall that  $\gamma_q$ controls overall
abundance of  quark pairs, while $\lambda_i$ control the difference
between quarks and anti-quarks of given flavor.
The difference between $\lambda_i$ and $\gamma_i$ is that, 
for quarks and anti-quarks the same factor $\gamma_{i}$ 
applies, while the antiparticle fugacity is inverse of the particle
fugacity. The phase space occupancy $\gamma_{q}$ 
is determined to exceed unity  significantly, once introduced:
results obtained switching on progressively more parameters
of the Fermi model show considerable improvement in the 
capability of the model to describe the data \cite{RLadv99}.
 
The proper statistical physics foundation of  $\gamma_i$ 
is obtained considering the maximum entropy principle: it has
been determined that while the limit $\gamma_i\to1$ maximizes the specific
chemical entropy, this maximum is not very  pronounced, 
indicating that 
a system with dynamically evolving volume 
will, in general, find more effective
 paths to increase entropy, than offered by the establishment of 
the absolute chemical equilibrium \cite{entro}. 

It is usual to introduce the geometric average of the light
quark fugacities, related to the baryochemical potential: 
\begin{equation}\label{fuga1}
\lambda_q^2\equiv \lambda_d\lambda_u\,,
\quad \mu_q\equiv T\ln \lambda_q\,,\quad \mu_B\equiv 3\mu_q\,.
\end{equation}
Similarly, one introduces the chemical 
potentials for up and down quarks:
\begin{equation}\label{fuga2}
\mu_d\equiv T\ln \lambda_d\,, \quad \mu_u\equiv T\ln \lambda_u\,,
\quad \delta\mu\equiv\mu_d-\mu_u\,.
\end{equation}
The difference in the number of net up ($u-\bar u$) and net down 
($d-\bar d$) quarks participating
in the dense matter fireball is given by the initial condition,
{\it i.e.}, which nuclei collide. In a hadron gas, 
$\delta\mu/\mu_q$  varies as function of chemical freeze-out 
temperature as shown in \cite[Fig.\,1]{Let95}.

 The relative number of  particles is controlled by
fugacity and phase space occupancies of the constituents.
The composite hadronic  particle chemical occupancy and 
fugacity is  expressed
by  constituent contributions:
\begin{equation}\label{abund2}
\lambda_i=\prod_{j\in i}\lambda_j\,,
\qquad  \gamma_i=\prod_{j\in i}\gamma_j\,.
\end{equation}
The abundances of particles produced in QGP disintegration is most 
conveniently described by considering the  phase space 
distribution of particles: the abundance of a hadron $h$, 
with $j$-components, freezing out from the source is
\begin{equation}\label{abund1}
N_h\propto \gamma_h e^{-E_h/T}\,,\qquad
\gamma_h\equiv\prod_{j\in h}\gamma_j\lambda_j\,.
\end{equation}

For example, the effective  
chemical fugacities for  $\pi^\mp$ composed of  light quark-antiquark pair
of a differing flavor is
\begin{equation}\label{gammamp}
 \gamma_{\pi^-}=\gamma_{q}^{2}\frac{\lambda_d}{\lambda_u}\,,\qquad
\gamma_{\pi^+}=\gamma_{q}^{2}\frac{\lambda_u}{\lambda_d}\,,
\end{equation}
while for neutral pions we simply have:
\begin{equation}\label{gamma0}
 \gamma_{\pi^0}=\gamma_{q}^{2}\,.
\end{equation}

The case of pions is, however, exceptional in another way:
the SPS data analysis shows  that the pion yield is governed by 
a large fugacity, and thus it is imperative to revert in the Fermi 
model to  use Bose distribution function: 
\begin{eqnarray}\label{piBos}
\frac {d^3N_{\pi}}{d^3p}=\frac{1}
    {\gamma_{\pi}^{-1}
        e^{E_\pi/T}-1}\,,\qquad
E_\pi=\sqrt{m_\pi^2+p^2}  \,.
\end{eqnarray}
We see that the range of values for $\gamma_{\pi}$ 
is bounded from above by the Bose singularity:
\begin{equation}\label{gammaqc}
\gamma_{\pi}< \gamma_{\pi}^c=e^{m_\pi/T}\,.
\end{equation}

When $\gamma_{\pi}\to \gamma_{\pi}^c$
the lowest energy state (in the continuum limit with $p\to 0$ )
will acquire macroscopic  occupation and a pion condensate is formed. 
Such a condensate `consumes' energy without
consuming entropy of the primordial high entropy QGP phase \cite{Let93,Let95}
and we cannot expect this to occur in hadronization of a high 
entropy phase into a low entropy confined phase. However, a sudden hadronization
process will naturally have the tendency to approach the limiting value 
$\gamma_{\pi}\to \gamma_{\pi}^c$ in
order to more efficiently connect the deconfined and 
the confined phases, since, as we show in Fig.\,\ref{abssne},
the entropy density is nearly twice as high at $\gamma_{\pi}\simeq 
\gamma_{\pi}^c$ than at $\gamma_\pi=1$. We note
that all thermodynamic quantities of interest here do not develop
a singularity in the limit $\gamma_{\pi}\to\gamma_{\pi}^c$.
\begin{figure}[tb]
\vspace*{-1.9cm}
\centerline{
\psfig{width=8.5cm,figure=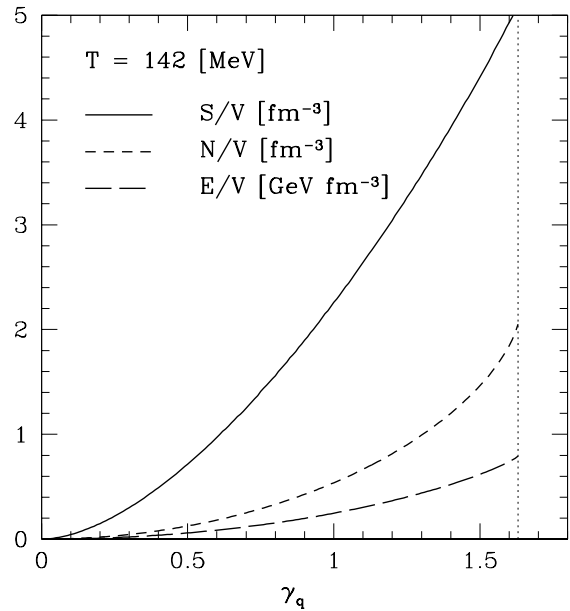}}
\vspace*{-1.8cm}
\caption{ 
Dependence of pion gas properties $N/V$-particle, $E/V$-energy and 
$S/V$-entropy  density, as function of  $\gamma_{q}$
(for $\lambda_d=\lambda_u$)  at $T=142$\,MeV. 
\label{abssne}} 
\end{figure}

We thus consider as a possible and indeed likely QGP
hadronization mechanism, the conversion of the excess of
entropy of the deconfined state 
into oversaturation of pion phase space. This mechanism
can replace the more conventional 
approaches that also described the `absorption' 
of the excess QGP entropy: \\
1) 
formation of a mixed phase which allows the volume of the
fireball of hadronic matter to grow in the hadronization; the
excess of QGP in entropy is converted  into the greater volume of lower 
entropy density hadronic gas. \\
2) 
reheating which allows the entropy excess to be accommodated in the
 momentum `volume' of the phase space.

Of course, both these effects can coexist, and also they can, in
principle, coexist with our proposed third mechanism, 
the chemical non-equilibrium. In fact, which exact path is taken
is only found in a microscopic description which allows for all 
these effects with appropriate relative rates. 
A model study of the dynamics of the hadronization process 
shows that the reaction mechanism here proposed is 
possible \cite{Bir99}, though a full understanding of the 
microscopic hadron phase transformation remains at present 
beyond our capability. 
 However, aside of the fact that the data analysis heavily
favors our proposal, we note that the conventional 
 processes are by far more complex and thus more difficult 
to realize on the short time scale ${\cal O}(10^{-22}\mbox{\ s})$
available in heavy ion collisions, as compared to a rather
straight forward gluon fragmentation, accompanied by recombination
of quarks into pions, which is the probable microscopic 
origin of the large chemical
nonequilibrium value of~$\gamma_q$.

Let us return now to the study of pion abundances: the slight 
chemical asymmetry originating in the larger neutron number than 
proton number in heavy nuclei, leads to light quark abundance
asymmetry, $N_d>N_u$, and requires that  one
of the charged pion fugacities  ($\pi^-$) increases (towards the condensation point), 
while  the other charged pion ($\pi^+$) fugacity is moved further away from it.
The result of this is clearly visible in Eq.\,(\ref{piBos})
when combined with Eqs.\,(\ref{gammamp},\ref{gamma0}):
as function of energy this slight change has the effect of enhancing the yield of 
low energy $\pi^-$ over that of $\pi^+$. There is a
much more significant effect when we are near to $\gamma_\pi^c$  than
it is the case for $\gamma_q=1$ which was explored earlier \cite{GM97}. 
In the following, we will address in detail the statistical parameters as 
they are determined for the 158--200$A$\,GeV interactions and 
presented in table \ref{param}. We note that for the Pb--Pb system, the value 
of $\lambda_{s}=1.10$ compensates the Coulomb effect distortion of the 
phase space of strange quarks in QGP. Thus, it is a measure of the magnitude of 
the Coulomb potential at hadronization, and hence of 
the freeze-out volume \cite{LRPb98}. We so estimate the volume of
the fireball at chemical freeze-out to be $V\simeq 2100~\mbox{fm}^3$, and this 
value is used to check if the absolute yields of pions, and in 
particular here negative hadrons ($\pi^-,$\,K$^-$,\,$\bar p$), 
are consistent with   measurement of the experiment NA49 \cite{App99}.
We find  within our approach that  among the 200 `negatives'  
at central rapidity $\Delta y=1$, 
there are  (roughly) about 145 directly produced $\pi^-$, 20
$\bar p$ and $K^-$, and only about 30 $\pi^-$ that arise from 
resonance decays. 

Fig.\,\ref{delpipi} presents a case study of the effect on the
thermal pion $m_\bot$ spectra. We integrate the longitudinal 
momenta in the range of rapidity $-0.5<y<0.5$ and consider the
difference in charged pion yield divided by the sum,
\begin{equation}\label{Dqmtpi}
D_q^\pi(m_\bot)\equiv \frac{N(\pi^-)-N(\pi^+)}{N(\pi^-)+N(\pi^+)}\,,
\end{equation}
as function of $m_\bot$, 
for small $m_\bot-m_\pi<0.3$\,GeV, showing the result on logarithmic
scale. The lowest  line shows the 
small effect of isospin asymmetry in Pb--Pb collisions when the
pion phase space is governed by the equilibrium chemical condition
$\gamma_q=1$; this result corresponds, for the range of our
statistical parameters here applicable, to the results
reported in Ref.\,\cite{GM97}.
As we move up,  for each of the lines in Fig.\,\ref{delpipi},
we increment $\gamma_q=1$ by 0.1 and the thick and top line
corresponds to the value implied by the analysis of the experimental
results, $\gamma_q=1.6$.
The magnitude of the effect is now quite surprising (remember 
logarithmic scale), and most importantly,
we do see a very clear sensitivity to the value of $\gamma_q$. While the
`modulator' of this sensitivity is the isospin asymmetry $\lambda_d/\lambda_u$,
this quantity is in principle measured by the results shown in Fig.\,\ref{delpipi},
since for  $m_\bot-m_\pi > 0.3$\,GeV we have:
\begin{equation}\label{Dqmtpi3}
D_q^\pi(m_\bot)\big|_{>0.3\,\mbox{\scriptsize GeV}}\simeq
 \frac{(\lambda_d/\lambda_u)^2-1}{(\lambda_d/\lambda_u)^2+1}\,.
\end{equation}
It must be stressed that the result shown in Fig.\,\ref{delpipi}
must be compared to experiment with some caution, since it
is just a case study aiming to show the extraordinary sensitivity  of the
$D_q^\pi(m_\bot)$ observable to the quantity $\gamma_q$. In order to 
be able to interpret actual data, one has to study the diluting pion component from
resonance decay and the effect of collective flow. 
It can be hoped that the flow effect
is small, and indeed  cancelling in the pion yield  ratio.
The contribution of resonance decays turns out also to be rather small in the
range of statistical parameters considered in view
of the small chemical freeze-out temperature.

\begin{figure}[tb]
\vspace*{-0.5cm}
\centerline{
\psfig{width=8.5cm,figure=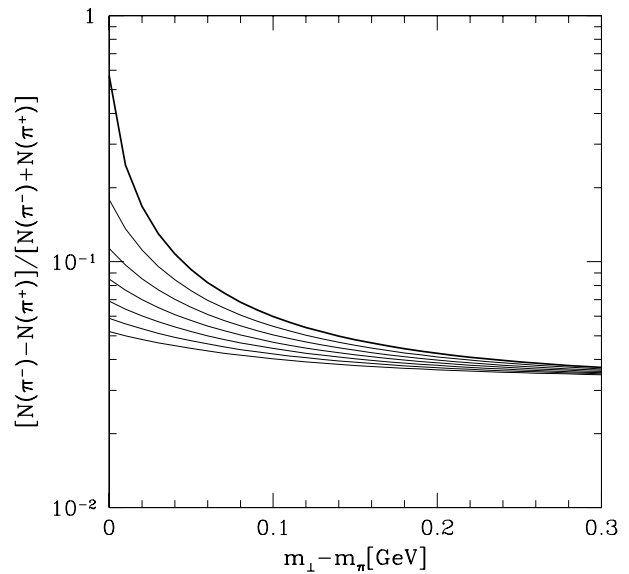}}
\vspace*{-0.2cm}
\caption{ 
Thermal pion asymmetry $m_\bot$ spectrum, for $\gamma_q=1\ \mbox{to}\ 1.6$ 
(thick line) by step of 0.1,
integrated in the rapidity window $-0.5<y<0.5$. 
\label{delpipi}} 
\end{figure}
\begin{table}[tb]    
\caption{\label{param} 
Chemical freeze-out statistical parameters for
Pb--Pb 158$A$\,GeV data \protect\cite{RLprl99} 
and S--Au/W/Pb 200$A$\,GeV data \protect\cite{LRa99}.
Last line is adapted from \protect\cite[Fig.\,1]{Let95}.}
\vspace{-0.2cm}\begin{center}
\begin{tabular}{lcc}
                       &  Pb                & S\\
\tableline
$T_{f}$ [MeV]          &    142 $\pm$ 3\ \ \      &  144 $\pm$ 2\ \ \  \\
$\lambda_{q}$          &   1.61 $\pm$ 0.02  &   1.51 $\pm$ 0.02 \\
$\lambda_{s}$          &   1.10 $\pm$ 0.02  & 1.00 $\pm$ 0.02   \\
$\gamma_{q}$           &   1.6 $\pm$ 0.5    &   1.41 $\pm$ 0.08 \\
$\gamma_{s}/\gamma_{q}$&   0.80 $\pm$ 0.05  &  0.69 $\pm$ 0.03  \\
$\delta\mu/\mu_q$       &   0.07             & 0.04   \\        
\end{tabular}
\end{center}
\vskip -0.8cm
\end{table}

Given the magnitude of the $\gamma_q$-effect, we also studied in Fig.\,\ref{pi+pi}
the spectrum of  neutral pions, which is practically indistinguishable 
from the average of the  charged pion  spectra, $(\pi^++\pi^-)/2$. 
The solid line shown in Fig.\,\ref{pi+pi} are not renormalized,
and (on logarithmic scale) the difference between 
the curves for $m_\bot-m_\pi>0.3$\,GeV
is the result of the varying yield coefficient $\gamma_q^2$.  More importantly, for  $m_\bot-m_\pi<0.2$\,GeV with growing  $\gamma_q$
we see a strong up-bending of the pion spectrum. 
Dashed lines in Fig.\,\ref{pi+pi} 
show results  for relatively small $\Delta y=0.02$ and 
demonstrate that there is just a minimal spectral deformation due
to a finite central $\Delta y=1$ integral. 

\begin{figure}[tb]
\vspace*{-0.7cm}
\centerline{
\psfig{width=8.5cm,figure=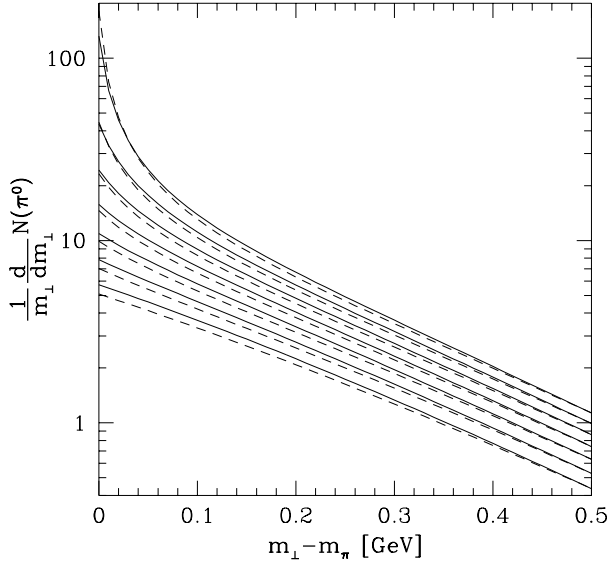}}
\vspace*{-0.2cm}
\caption{ 
Thermal $\pi^0$  spectrum, $\frac{1}{m_\bot}\frac{dN(\pi^0)}{dm_\bot}$, for 
$\gamma_q=1\ \mbox{to}\ 1.6$ by step of 0.1, 
integrated in the rapidity window: $-0.5<y<0.5$ (full lines),
and $-0.01<y<0.01$ (dashed lines), the latter normalized to the
result with $\Delta y=1$ at $m_\bot-m_\pi=0.5$\,GeV.
\label{pi+pi}
} 
\end{figure}

In order to understand the experimental pion spectra
it is necessary to consider the $\gamma_q$-effect, 
combined with  the flow effect and the hadron disintegration
feeding of pion spectra.
We recall here that aside of direct influence on primordial 
pions, $\gamma_q$ also impacts the relative yield of primordial 
mesons compared to the yield of secondary mesons from resonance decays.
Thus omission to consider  $\gamma_q$-effect can mislead the
data interpretation.  
We recall here a recent analysis of the $\pi^0$  spectra which has
reached somewhat unorthodox conclusions about the freeze-out 
conditions as seen  in Ref.\,\cite[Table\,1]{WA98freeze}. 
We believe that allowance for chemical  non-equilibrium, 
specifically the introduction of $\gamma_q\simeq \gamma_q^c$ 
should alter the conclusions of the WA98 collaboration. Even though they
study the pion spectrum at rather large $m_\bot$,  $\gamma_q$
impacts the cocktail mix of  direct with decay pions.
There is another detailed fit 
which also restricted the single particle pion spectra
to be exactly at chemical equilibrium\cite{HeiF99}. 
However, these authors do note that in order to describe the 
pion yield, they need a pion fugacity 
$\gamma_q^{100}=1.36$, a value which may (barely) allow their 
tacit assumption $\gamma_q(T=100\,\mbox{MeV})=1$, 
made when they interpret particle spectra.  
At the higher  value of chemical freeze-out 
temperature a range of $\gamma_q$ is expected that 
in our opinion requires reevaluation  of their results. 
 On the other hand, the parallel analysis of Schlei and
collaborators \cite{Sch99}, which did not attempt a precise fit, and did 
not describe the low $m_\bot$ pion spectra, and which obtains a similar
chemical freeze-out temperature as used here, appears to us to remain
valid.

\begin{figure}[tb]
\vspace*{-1.cm}
\centerline{\hspace*{-1cm}
\psfig{width=8.5cm,figure=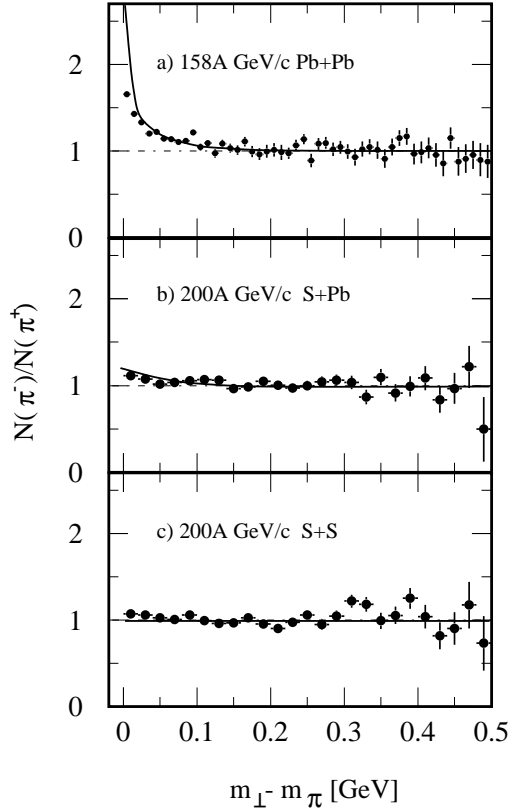}}
\vspace*{-1.7cm}
\caption{ 
Central rapidity charged pion ratio as function of $m_\bot$: data are by
NA44 collaboration \protect\cite{NA44C}, arbitrarily normalized to unity at 
high $m_\bot$. The lines are our result for $\gamma_q=1.6$, also renormalized to
unity at large $m_\bot$.
\label{pi/pi}} 
\end{figure}

It would be interesting to see what effect would arise
after explicit allowance for a  pion fugacity is made
in these studies. Since we do not model here the effect 
of collective flow as was 
done in Refs.\,\cite{WA98freeze,HeiF99,Sch99}, we cannot ourselves 
answer this question in full. However, 
there is data from NA44 on charged pion ratios\cite{NA44C}. 
Considering:\\
1) the relatively small dilution of primordial pion 
spectra by pions from resonance decays,
expected for the range of chemical freeze-out 
parameters shown in table \ref{param}, and \\
2) the  cancellation of the flow deformation of the spectra in spectral
ratios of particles of the same mass,\\ 
we have evaluated  the expected $\gamma_q$-effect for the charged pion ratio 
results presented
by NA44 collaboration \cite{NA44C}, as shown  in Fig.\,\ref{pi/pi}.
We note that we slightly overpredict at small $m_\bot$ these results, an effect
that is required given the inclusion of hadron resonance decay feeding into
the  experimental data. 

Several attempts have been made before to explain these NA44 data.
The experimental group presented calculational results obtained with
RQMD(v1.08) \cite{NA44C}, which could not account for the data.  
Even though NA44 considered in their numerical model also hadron 
resonance  decay effect, a scheme has been proposed 
to create the observed pion asymmetry from strangeness 
asymmetry \cite{Arb97}: in presence of finite baryon number 
there are more $s$-carrier  baryons than antibaryons and
more $\bar s$-carrier mesons than $s$-carrier mesons. Of course, such 
a description can easily be falsified by study of strange hadron abundances.
In thermal models, for $\gamma_q=1$, the NA44 asymmetry cannot be explained 
for the widest  range of reasonable statistical parameters, a fact we 
have realized long ago, and which is implicit in other work \cite{GM97}.
These studies have thus lead to the believe that 
Coulomb field effects in single  particle 
spectra are the source of pion asymmetry and a large number
of authors has explored this hypothesis \cite{Osa97,Bar98,Aya99,OH99}.  
It is quite possible that in part the effect of the 
pion asymmetry is due to the Coulomb effect, and in part explained by the 
$\gamma_q$-effect here described, considering that the
value $\gamma_q=1.6$ obtained from the study of hadron abundances has a 
significant statistical error. 

The Coulomb effect work is based on the picture of charged pions
acted upon by  the Coulomb field after   formation at a 
classically well determined point. The pion asymmetry is a
result of the  hypothesis that the  pion formation weight $W^\pm$ is  
actually equal for both  $\pi^+$ and $\pi^-$  at equal local momentum: 
\begin{equation}\label{WKBpiW}
\frac{d^3W^\pm}{d^3p}\propto \frac{1}{e^{\sqrt{p(R_n)^2+m_\pi^2}/T}-1}\,,
\end{equation}
where $R_n$ is a static or dynamic formation surface. 
In the WKB approximation:
\begin{equation}\label{WKBpi}
p^\mp(x)=\sqrt{(E_\pi\mp V(x))^2-m_\pi^2}\,.
\end{equation}
We believe that considerable improvement in these studies of the 
Coulomb effect is possible. In our view,
 the  Coulomb field influences the quantum density of 
states obtainable from the scattering phase $\eta$. This introduces 
in addition to  the statistical density 
matrix weight $\mbox{Tr}\,e^{-\beta H}$ 
an additional  density of states weights 
$d\eta^{\pi^-}\!\!\!/dE,\ d\eta^{\pi^+}\!\!\!/dE$,   
which  due to the presence of attractive 
(for $\pi^-$) and repulsive (for $\pi^+$)
Coulomb potential is Coulomb deformed at 
asymptotic pion momenta values that sample 
the physical  size of the potential. 
We note that for $R_n=8$--$10$\,fm
this effect is expected to occur for 
$E_\pi-m_\pi<\pi^2/2R_n^2m_\pi\simeq 10$--$20$\,MeV in static case. 
In quantum formulation the presence of (Coulomb) potential alters locally the 
reduced wave  length  $\lambda^{\!\!\!\!-}=\hbar/p_\pi(x)$,  but not the 
asymptotic energy $E_\pi$ of a  quantum state. 
The work  most closely related to this approach \cite{Osa97,OH99}
takes Coulomb wave-functions 
(thus better than the  WKB illustration made above) but adds ad hoc  
density of state hypothesis,
see Ref\,\cite[Eq.\,(2)]{Osa97}, using here the density of matter
in configuration space folded with the Coulomb wave. 
The pion asymmetry is arising as result of substitution of Eq.\,(\ref{WKBpi})
in  Eq.\,(\ref{WKBpiW}).  

Our findings have  interesting bearing on the 
forthcoming experimental RHIC results. The data from the 
experiment PHOBOS \cite{Phobos}
should in our opinion also allow to observe the 
 $\gamma_q$-effect, both in soft pion spectra, 
and in pion asymmetry, provided that 
the non-equilibrium occupancy, $\gamma_q\simeq 1.6$, which we
associated with the QGP hadronization, also occurs  at RHIC --- 
as we expect \cite{RL99}. The analysis of these data should follow
the pattern outlined here: from the slight inequality of the 
abundance at moderate $m_\bot$, according to Eq.\,(\ref{Dqmtpi3}), 
the value of chemical asymmetry can be deduced and this is 
used to determine the value of $\gamma_q$, given the shape of the 
low-$m_\bot$ spectral asymmetry following from Eq.\,(\ref{piBos}). 
We note that  Coulomb  analysis also predicts a similar effect 
at RHIC energies  \cite{OH99}.

We have shown that the phase space overoccupancy
$\gamma_q\simeq 1.6$ of pions arising  in the sudden 
hadronization process of the deconfined quark-gluon 
plasma enhances the low-$m_\bot$ pion spectra, and amplifies the 
abundance ratio of charged pions.  We have quantitatively 
demonstrated the great sensitivity of these effects to the value of 
 $\gamma_q>1$ and have shown how this $\gamma_q$-effect helps 
understand SPS soft pion results of NA44 and WA98 collaborations.
We conclude that the reported excess of soft pions and the
asymmetry of soft charged pion spectra provides evidence for
sudden and explosive  hadronization (glue fragmentation)
 of the entropy rich deconfined phase. 
This finding about the mechanism of 
hadronization  corroborates the conclusions we
 obtained studying high $m_\bot$ strange
hadron spectra. 

{\vspace{0.5cm}\noindent\it Acknowledgments:\\}
Supported in part by a grant from the U.S. Department of
Energy,  DE-FG03-95ER40937\,. LPTHE, Univ.\,Paris 6 et 7 is:
Unit\'e mixte de Recherche du CNRS, UMR7589.\vskip -0.5cm


\end{narrowtext}

\end{document}